\documentclass[aps,twocolumn,superscriptaddress]{revtex4-1}
\usepackage{amsmath,amssymb}
\usepackage{slashed}
\usepackage{graphicx}
\usepackage{bm}
\usepackage[T1]{fontenc}
\usepackage[utf8]{inputenc}
\usepackage{gauss} 
\usepackage[top=1.0in,bottom=1.0in,left=1.0in,right=1.0in]{geometry}
\usepackage[colorlinks,linkcolor=blue,citecolor=blue]{hyperref}
\usepackage{comment}
\usepackage{multirow}
\usepackage{ytableau}
\usepackage{makecell}
\usepackage{caption}
\usepackage{tabularx}
\usepackage{float}
\newcommand{\be}{\begin{equation}}
\newcommand{\ee}{\end{equation}}
\newcommand{\bea}{\begin{eqnarray}}
\newcommand{\eea}{\end{eqnarray}}
\newcommand{\ba}{\begin{eqnarray}}
\newcommand{\ea}{\end{eqnarray}}

\begin{document}

\title{Light S-wave pentaquarks  on the light front}

\author{Fangcheng He}
\email{fangcheng.he@stonybrook.edu}
\affiliation{Center for Nuclear Theory, Department of Physics and Astronomy, Stony Brook University, Stony Brook, New York 11794--3800, USA}

\author{Edward Shuryak}
\email{edward.shuryak@stonybrook.edu}
\affiliation{Center for Nuclear Theory, Department of Physics and Astronomy, Stony Brook University, Stony Brook, New York 11794--3800, USA}

\author{Wan Wu}
\email{wuw20@mails.tsinghua.edu.cn}
\affiliation{Center for Nuclear Theory, Department of Physics and Astronomy, Stony Brook University, Stony Brook, New York 11794--3800, USA}
\affiliation{Physics Department, Tsinghua University, Beijing 100084, China}

\author{Ismail Zahed}
\email{ismail.zahed@stonybrook.edu}
\affiliation{Center for Nuclear Theory, Department of Physics and Astronomy, Stony Brook University, Stony Brook, New York 11794--3800, USA}

\begin{abstract} 
We construct an explicit basis set for pentaquark states on a regular 4-simplex, that diagonalizes the Hamiltonian for light pentaquarks with confinement on the light front (LF). The ensuing eigenstates are free of the center of mass motion and satisfy  exact Dirichlet boundary conditions. Hyperfine interactions in the form of color-spin or flavor-spin are shown to lift the degeneracy of the 16 pentastates, with a spectrum that compares fairly with some of the empirical nucleon excited states. The quark PDF for the light pentastates is discussed.
 \end{abstract}
\maketitle

\section{Introduction}
Constituent quark models emerge from our current understanding of the QCD vacuum at low resolution, with running constituent masses and effective couplings with spontaneously
broken chiral symmetry and bound light hadrons. They yield to a modern understanding of current spectroscopy in the center of mass frame. 

To address issues related to the partonic constituents of hadrons, a formulation on the light front is needed. In the latter,
light and heavy degrees of freedom are treated democratically, both in terms of straight Wilson lines thanks to the infinite momentum frame. In addition what is a cloud of mesons in the rest frame, translates to a parton sea by boosting, in the form of a multi-Fock component
of a hadron. 

For a baryon with 3-quarks the simplest Fock extension is a pentaquark state. In a recent
analysis~\cite{Miesch:2025wro}, two of us have shown that the S- and P-shell pentaquark states mix with the standard 3-quark states through induced scalar and pseudo-scalar interactions (meson clouds), in the center of mass (CM) frame. This mixing is central for the explanation of the emergence of a large orbital contribution to the baryon angular momentum, and the strong d-u flavor asymmetry  in nucleons. 

The aim of this paper is to construct S-wave pentaquark states by  diagonalization of the 5-quark light cone Hamiltonian, which is free of the CM motion. In contrast to other approaches~\cite{Wiecki:2014ola}, 
where the center of mass motion is present in the lightcone eigenbasis but removed only  by projection through a Lagrange multiplier proportional to CM motion added  to the Hamiltonian, our construction of the longitudinal and transverse eigenbases is inherently free of CM motion.
In particular, the longitudinal eigenbasis satisfies the exact Dirichlet boundary conditions on a regular pentachoron or a 4-simplex, appropriate for a five-quark state.
This construction, complements our recent results using variational analyses~\cite{Miesch1}. 

The paper is organized as follows. In section~\ref{sec_II} we develop the bare
LF Hamiltonian for light pentaquarks in Jacobi coordinates, after removing the CM motion. In section~\ref{sec_III} we detail the construction of the eigenbasis on  a simplex borrowing  the exact solutions to identical and impenetrable fermions on a line. The pentaquark states are then constructed by direct
diagonalization of their pertinent LF Hamiltonian.
In section~\ref{sec_V} the hyperfine
splitting of the S-states pentaquarks are worked out for total antisymmetrization of the color-spin-flavor states. The color-spin and flavor-spin splittings yield light pentastates 
which are comparable to the measured nucleon states. The numerical results for the ground state eigenfunctions and hyperfine
splitting of s-wave states are presented in section~\ref{sec:numres}. 
Our conclusions are in section~\ref{sec_VI}.

\section{LF Hamiltonian and Jacobi coordinates} 
\label{sec_II}
Our description of the light pentaquarks $qqqq\bar q$ will follow that of the light baryons
$qqq$ on the LF~\cite{Shuryak:2022thi}. The light front Hamiltonian for the pentaquark state includes the kinetic and confining terms
\bea
H'_{LF}&=&\sum_{i=1}^5\Bigg(\frac{k^2_{i\perp}+m_Q^2}{x_i}
\nonumber\\
&+&2\sigma_T((i\partial/\partial x_i)^2+M^2 r^2_{i\perp})^{1/2}\Bigg)
\eea
where $\sigma_T$ is the string tension and (in lowest order) $M\approx 5m_Q$. Here $x_i$ are momentum fractions
for quarks, $i=1..5$ and $r_{i\perp}$ are transverse coordinates.

The momentum conservation conditions for the pentaquark state are $\sum_{i=1}^5\vec{k}_{i\perp}=0$ for transverse momenta and $\sum_{i=1}^5x_i=1$ for longitudinal momentum fraction, $0\leq x_i\leq 1$. 
Using the einbein trick, one can get rid of the square roots in the conﬁning term,
\bea\label{eq:LFHami}
H'_{LF}&=&\sum_{i=1}^5\frac{k^2_{i\perp}+m_Q^2}{x_i}
\nonumber\\
&+&\sigma_T\left(5a+\frac{1}{a}\sum_{i=1}^5((i\partial/\partial x_i)^2-M^2 (\partial/\partial_{\vec{k}_{i\perp}})^2)\right) \nonumber\\
\eea
where the transverse coordinates have been replaced by $r^2_{i\perp}=-\partial/\partial_{k_{i\perp}^2}$.
Following~\cite{Shuryak:2022thi}, the symmetric einbein parameter $a$ will be fixed below by minimization of the resulting masses, with a value that depends on the constituent quark mass. 

To eliminate the motion of center mass, we use the following Jacobi coordinates for the longitudinal part,
\bea
x_\alpha&=&\frac{x_1-x_2}{\sqrt{2}},\nonumber\\
x_\beta&=&\frac{x_1+x_2-2x_3}{\sqrt{6}} ,\nonumber\\
x_\gamma&=&\frac{x_1+x_2+x_3-3x_4}{\sqrt{12}} ,\nonumber\\
x_\delta&=&\frac{x_1+x_2+x_3+x_4-4x_5}{\sqrt{20}}
\eea
One can transfer the light cone coordinates $x_i$ to the Jacobi coordinates using the above equations with the momentum conservation condition $\sum_{i=1}^5x_i=1$,
\bea\label{eq:xtoJacb}
x_1&=&\frac{1}{30} \left(15 \sqrt{2} x_{\alpha }+5 \sqrt{6} x_{\beta }+3 \sqrt{5} x_{\delta }+5 \sqrt{3} x_{\gamma }\right)+\frac{1}{5},\nonumber\\
x_2&=&\frac{1}{30} \left(-15 \sqrt{2} x_{\alpha }+5 \sqrt{6} x_{\beta }+3 \sqrt{5} x_{\delta }+5 \sqrt{3} x_{\gamma }\right)+\frac{1}{5} ,\nonumber\\
x_3&=&\frac{1}{30} \left(-10 \sqrt{6} x_{\beta }+3 \sqrt{5} x_{\delta }+5 \sqrt{3} x_{\gamma }\right)+\frac{1}{5} ,\nonumber\\
x_4&=&\frac{1}{10} \left(\sqrt{5} x_{\delta }-5 \sqrt{3} x_{\gamma }\right)+\frac{1}{5},\nonumber\\
x_5&=&-\frac{2 \sqrt{5}}{5} x_{\delta }+\frac{1}{5}
\eea

The transverse part can also be similarly rewritten into the Jacobi coordinates, after enforcing momentum conservation $\sum_{i=1}^5\vec{k}_{i\perp}=0$. With the help of the einbein transform, the square-root in the confining potential (\ref{eq:LFHami}) can be unwound
\bea\label{eq:laplacian}
&&(i\partial/\partial x_i)^2+M^2 r^2_{i\perp})^{1/2}  \nonumber\\
&\rightarrow&-\left(\frac{\partial^2}{\partial x^2_\alpha}
+\frac{\partial^2}{\partial x^2_\beta}
+\frac{\partial^2}{\partial x^2_\gamma}
+\frac{\partial^2}{\partial x^2_\delta}\right)\nonumber\\
&-&M^2\left(\frac{\partial^2}{\partial \vec{k}^2_\alpha}
+\frac{\partial^2}{\partial \vec{k}^2_\beta}
+\frac{\partial^2}{\partial \vec{k}^2_\gamma}
+\frac{\partial^2}{\partial \vec{k}^2_\delta}\right)
\nonumber\\
&=&-\nabla_L-M^2\nabla_T
\eea
with the CM contribution dropped. The final expression for the light front Hamiltonian, free of CM, is then given by
\bea\label{eq:LFHamiJob}
H_{LF}&\equiv&\sum_{i=1}^5\frac{k^2_{i\perp}+m_Q^2}{x_i} + 5\sigma_Ta 
\nonumber\\
&-&\frac{\sigma_T}{a}\sum_{\xi=\alpha,\beta,\gamma,\delta}((\partial/\partial x_\xi)^2+M^2 (\partial/\partial_{\vec{k}_{\xi\perp}})^2) \nonumber\\
\eea

\section{Construction of an eigenbasis } 
\label{sec_III}
To diagonalize the Light front Hamiltonian in Eq.~(\ref{eq:LFHamiJob}), one needs to find the proper basis. For the longitudinal part, the momentum constraints
\bea
\label{SNCANO}
X=\sum_{i=1}^Nx_i=1 \qquad{\rm with}\qquad 1 \geq x_i\geq 0
\eea
For $N=3$ (the baryons) the   enclosed domain is equilateral triangle, and for $N=5$ (pentaquarks) it is a regular 4-dimensional simplex known as $pentachoron$, or ``standard simplex"  
$\Delta_N[x_1,...,x_N]$ defined by the Jacobi coordinates.
It has 5 ``corners" and 10 ``faces". 
Note that the WFs should vanish on all its faces (Dirichlet boundary conditions), which corresponds to some parton momentum vanishing $x_i=0$, due to the singular terms proportional to $1/x_i$ in the Hamiltonian (\ref{eq:LFHamiJob}).  

\subsection{General case for for longitudinal momenta}
We now note that the standard simplex 
$\Delta_N[x_1,...,x_N]$ we need to study is geometrically isomorphic  to the ``ordered simplex" $\Delta^*_N[s_1,...,s_N]$ in $R^N$
\bea
\label{SNORD}
 s_1\leq s_2\leq  ...\leq s_{N}\leq s_1+a
\eea
where parameter $a$ is fixed by the length of the side of simplex. The faces of the ordered simplex occur when two successive coordinates are equal $s_i=s_{i+1}$.
The LFWFs are then amenable to the WFs of 
$N$ particles (fermions) moving on a circle of unit circumference
\bea
\varphi[x]\rightarrow \varphi[s]\qquad {\rm with}\qquad
\varphi[s|s_i=s_j]=0
\eea
These $N$ impenetrable particles map onto $N$ 
free fermions,
thanks to the Pauli exclusion principle. The N-fermion
WFs are Slater determinants composed of circular waves~\cite{Girardeau:1960cnk}
\bea
\label{SLAT1}
\varphi_n[s]=\frac {1}{N^{4/3}}
\begin{vmatrix}
    e^{i\tilde n_1s_1}& e^{i\tilde n_1s_2}  & \dots  & e^{i\tilde n_1s_N} \\
    e^{i\tilde n_2s_1}& e^{i\tilde n_2s_2}  & \dots  & e^{i\tilde n_2s_N} \\
    \vdots & \vdots & \ddots & \vdots &  \\
      e^{i\tilde n_Ns_1}& e^{i\tilde n_Ns_2}  & \dots  & e^{i\tilde n_Ns_N} 
\end{vmatrix}\nonumber\\
\eea
with $\tilde n_i=2\pi n_i$. 
They have support on disjoint simplexes under permutation,  
with  eigen-energies
\be
E[n]=\sum_{i=1}^N\tilde n_i^2
\ee
A similar approach
was used to diagonalize the  free fermion problem with Dirichlet boundary conditions, on an irregular tetrahedron ~\cite{krishnamurthy1982exact}.

The ordering of the coordinates 
 $s_i\in [0,1]$,  suggests their identification with the coordinates of equidistant particles on a string 
of length 1 in 1-dimension,  undergoing longitudinal vibrations. With this in mind and
to eliminate the CM of mass coordinate, we shift $s_j=\frac{S}{N}+\tilde s_j$ with $S=\sum_{j=1}^N s_j$ in  (\ref{SLAT1}), and using the rearrangement of the rows and columns in the Slater determinant, we obtain
\begin{widetext}
\bea
\begin{vmatrix}
    e^{i\tilde n_1s_1}& e^{i\tilde n_1s_2}  & \dots  & e^{i\tilde n_1s_N} \\
    e^{i\tilde n_2s_1}& e^{i\tilde n_2s_2}  & \dots  & e^{i\tilde n_2s_N} \\
    \vdots & \vdots & \ddots & \vdots &  \\
      e^{i\tilde n_Ns_1}& e^{i\tilde n_Ns_2}  & \dots  & e^{i\tilde n_Ns_N} 
\end{vmatrix}=
\begin{vmatrix}
    e^{i\tilde n_1 \frac SN}e^{i\tilde n_1\tilde s_1}&  e^{i\tilde n_1 \frac SN}e^{i\tilde n_1\tilde s_2}  & \dots  &  e^{i\tilde n_1 \frac SN}e^{i\tilde n_1\tilde s_N} \\
     e^{i\tilde n_2 \frac SN}e^{i\tilde n_1\tilde s_1}e^{i\tilde n_{21}\tilde s_1}&  e^{i\tilde n_2 \frac SN} e^{i\tilde n_1\tilde s_2} e^{i\tilde n_{21}\tilde s_2}  & \dots  & e^{i\tilde n_2 \frac SN} e^{i\tilde n_1\tilde s_N}e^{i\tilde n_{21}\tilde s_N} \\
    \vdots & \vdots & \ddots & \vdots &  \\
      e^{i\tilde n_N \frac SN} e^{i\tilde n_1\tilde s_1}e^{i\tilde n_{N1}\tilde s_1}& e^{i\tilde n_N \frac SN} e^{i\tilde n_1\tilde s_2} e^{i\tilde n_{N1}\tilde s_2}  & \dots  &  e^{i\tilde n_N \frac SN}e^{i\tilde n_1\tilde s_N} e^{i\tilde n_{N1}\tilde s_N} 
\end{vmatrix}
\eea 
and  pulling common factors with the center of mass contribution, 
\bea \label{eqn_slater_N}
 e^{i\tilde n_1 \frac SN}
  e^{i\tilde n_2 \frac SN}...
   e^{i\tilde n_N\frac SN}
 e^{i\tilde n_1\tilde s_1}e^{i\tilde n_1\tilde s_2}...e^{i\tilde n_1\tilde s_N}
 =e^{i\frac SN\sum_{i=1}^N \tilde n_j}
\eea
yield the result
\bea
\label{SLAT4}
\varphi_n[s]=\frac {1}{N^{4/3}}
e^{i\frac SN\sum_{i=1}^N \tilde n_j}
\begin{vmatrix}
   1& 1  & \dots  & 1\\
    e^{i\tilde  n_{21}\tilde s_1}& e^{i\tilde  n_{21}\tilde s_2}  & \dots  & e^{i\tilde  n_{21}\tilde s_N} \\
    \vdots & \vdots & \ddots & \vdots &  \\
      e^{i\tilde  n_{N1}\tilde s_1}& e^{i\tilde  n_{N1}\tilde s_2}  & \dots  & e^{i\tilde  n_{N1}\tilde s_N} 
\end{vmatrix}
\equiv e^{i\frac SN{\sum_{i=1}^N \tilde n_j}}\varphi_{\tilde n}[\tilde s]
\eea
\end{widetext}
 with $\tilde n_{i1}=\tilde n_i-\tilde n_1$ for $i=2, .., N$.  The coefficients $\tilde n_{i1}$ are ordered as $0<\tilde n_{21}<\tilde n_{31}<\tilde n_{41}...$ to avoid repeated counting of the same state. One can get rid of the center mass and get the eigenvalues for the reduced WFs $\varphi[\tilde s]$, which are only a function of 
the $N-1$ coordinates, with energies
\bea
\label{SPECN}
\tilde E[\tilde n]&=&\frac{1}{\varphi_{\tilde{n}}[\tilde{s}]}\left[\left(\frac{\partial}{\partial s_i}\right)^2-N\left(\frac{\partial}{\partial S}\right)^2\right]\varphi_{\tilde{n}}[\tilde{s}]
\nonumber\\
&=&\sum_{i=1}^{N}\tilde n_i^2-\frac 1N\bigg(\sum_{i=1}^{N} \tilde n_i\bigg)^2 \nonumber\\
&=&\sum_{i=2}^{N}\tilde n_{i1}^2-\frac 1N\bigg(\sum_{i=2}^{N} \tilde n_{i1}\bigg)^2
\eea
Since the set of Slater determinants is orthonormal and complete, the solutions $\varphi_{\tilde n}[\tilde s]$ map onto a complete basis set. The normal mode coordinates   can be related to the Jacobi coordinates.

\subsection{Longitudinal basis for N=5}
Specifically, we are interested in the Pentaquark states in this study. The orthonormal basis can be expressed as
\bea
\label{SLAT4}
\varphi_{\tilde{n}}[\tilde{s}]=\frac {1}{5^{4/3}}
\begin{vmatrix}
   1& 1  & \dots  & 1\\
    e^{i\tilde  n_{21}\tilde s_1}& e^{i\tilde  n_{21}\tilde s_2}  & \dots  & e^{i\tilde  n_{21}\tilde s_5} \\
    \vdots & \vdots & \ddots & \vdots &  \\
      e^{i\tilde  n_{51}\tilde s_1}& e^{i\tilde  n_{51}\tilde s_2}  & \dots  & e^{i\tilde  n_{51}\tilde s_5} 
\end{vmatrix}
\eea
To enforce the Dirichlet boundary conditions on the pentachoron formed by the longitudinal momenta
\bea
\label{5SIMPLEX}
\sum_{i=1}^5x_i=1 \qquad\qquad 0\leq x_i\leq 1
\eea
and keep the determinantal form
of the solutions, we construct the linear map
\bea
\label{SX}
\tilde{s}_1-\tilde{s}_2&=&s_1-s_2=x_1\geq 0\nonumber\\
\tilde{s}_2-\tilde{s}_3&=&s_2-s_3=x_2\geq 0\nonumber\\
\tilde{s}_3-\tilde{s}_4&=&s_3-s_4=x_3\geq 0\nonumber\\
\tilde{s}_4-\tilde{s}_5&=&s_4-s_5=x_4\geq 0
\eea
expression Jacobi coordinates with differences of circle coordinates. The above conditions ensure that the eigenbasis defined in Eq.~(\ref{SLAT4}) vanishes at $x_i = 0$. Note that those are subject to the CM constraint $$\sum_{i=1}^5\tilde{s}_i=0,~~~~\sum_{i=1}^5s_i=S$$ which preserves the ordering. 
Note that the left out coordinate $x_5$ is fixed by the longitudinal momentum constraint (\ref{5SIMPLEX}),  with $x_{1,..,4}$  defined in terms of Jacobi variables defined in Eq~(\ref{eq:xtoJacb})
 we can express $\tilde{s}_i$ and $s_i$ in (\ref{SX}) using Jacobi variables as
\begin{widetext}    
\bea
\label{MAPS}
s_1&=&\tilde{s}_1+\frac{S}{5}=\frac{1}{10} \left(2\sqrt{5} x_{\delta }+2\sqrt{3} x_{\gamma }+4+\sqrt{2} x_{\alpha }+\sqrt{6} x_{\beta }\right)+\frac{S}{5}
\nonumber\\
s_2&=&\tilde{s}_2+\frac{S}{5}=\frac{1}{30} \left(-12 \sqrt{2} x_{\alpha }-2 \sqrt{6} x_{\beta }+3 \sqrt{5} x_{\delta }+\sqrt{3} x_{\gamma }+6\right)+\frac{S}{5}
\nonumber\\
s_3&=&\tilde{s}_3+\frac{S}{5}=\frac{1}{30} \left(3 \sqrt{2} x_{\alpha }-7 \sqrt{6} x_{\beta }-4 \sqrt{3} x_{\gamma }\right)+\frac{S}{5}
\nonumber\\
s_4&=&\tilde{s}_4+\frac{S}{5}=\frac{1}{10} \left(\sqrt{2} x_{\alpha }+\sqrt{6} x_{\beta }-\sqrt{5} x_{\delta }-3 \sqrt{3} x_{\gamma }-2\right)+\frac{S}{5}
\nonumber\\
s_5&=&\tilde{s}_5+\frac{S}{5}=\frac{1}{10} \left(-2\sqrt{5} x_{\delta }+2\sqrt{3} x_{\gamma }-4+\sqrt{2} x_{\alpha }+\sqrt{6} x_{\beta }\right)+\frac{S}{5}\nonumber\\
\eea
\end{widetext}

The basis in Eq.~(\ref{SLAT4}) can be constructed directly from the $\tilde{s}_i$ defined above. Moreover, the Laplacian now takes the form of the following elliptic operator
\bea
\label{ELLIPTIC5}
\sum_{i=1}^5\partial^2_{s_i}&=&5\partial_{S}^2+\frac 16
\bigg(18\partial^2_{x_\alpha}+14\partial^2_{x_\beta}+13\partial^2_{x_\gamma}+15\partial^2_{x_\delta}\nonumber\\
&&-4\sqrt 3 \partial_{x_\alpha}\partial_{x_\beta}+\sqrt 6\partial_{x_\alpha}\partial_{x_\gamma}+3\sqrt{10}\partial_{x_\alpha}\partial_{x_\delta}\nonumber\\
&&-7\sqrt 2 \partial_{x_\beta}\partial_{x_\gamma}
+\sqrt{30} \partial_{x_\beta}\partial_{x_\delta}
-2\sqrt{15}\partial_{x_\gamma}\partial_{x_\delta}\bigg)\nonumber\\
&=&5\partial_{S}^2+\partial_X^2
\eea
The coordinate map (\ref{MAPS}) when used in  (\ref{SLAT4}), yields a complete and orthonormal set of Slater determinants that vanish on the 10 equilateral faces of the pentachoron as defined by the 4-simplex (\ref{5SIMPLEX}), solely in terms of the Jacobi coordinates $x_\alpha, x_\beta, x_\gamma, x_\delta$. The eigenvalues of $\partial_X^2$ are consistent with those given in~(\ref{SPECN}).

The resulting set of Slater determinants 
can be used as a basis set to diagonalize the LC Hamiltonian where the Laplacian is now a general elliptic operator (\ref{ELLIPTIC5}) plus a potential term. 
Note that the general elliptic operator becomes the standard Laplacian operator when $N=3$, the eigenvectors obtained using this method are consistent with the ones used in~\cite{Shuryak:2022thi}. For the $N=5$ case it is not, but one can still use the basis defined in Eq.~(\ref{SLAT4}) to diagonalize the Laplacian operator with the Dirichlet
boundary condition on a regular 4-simplex. 

\begin{figure}
\begin{center}
\includegraphics[width=7cm, height=5.5cm]{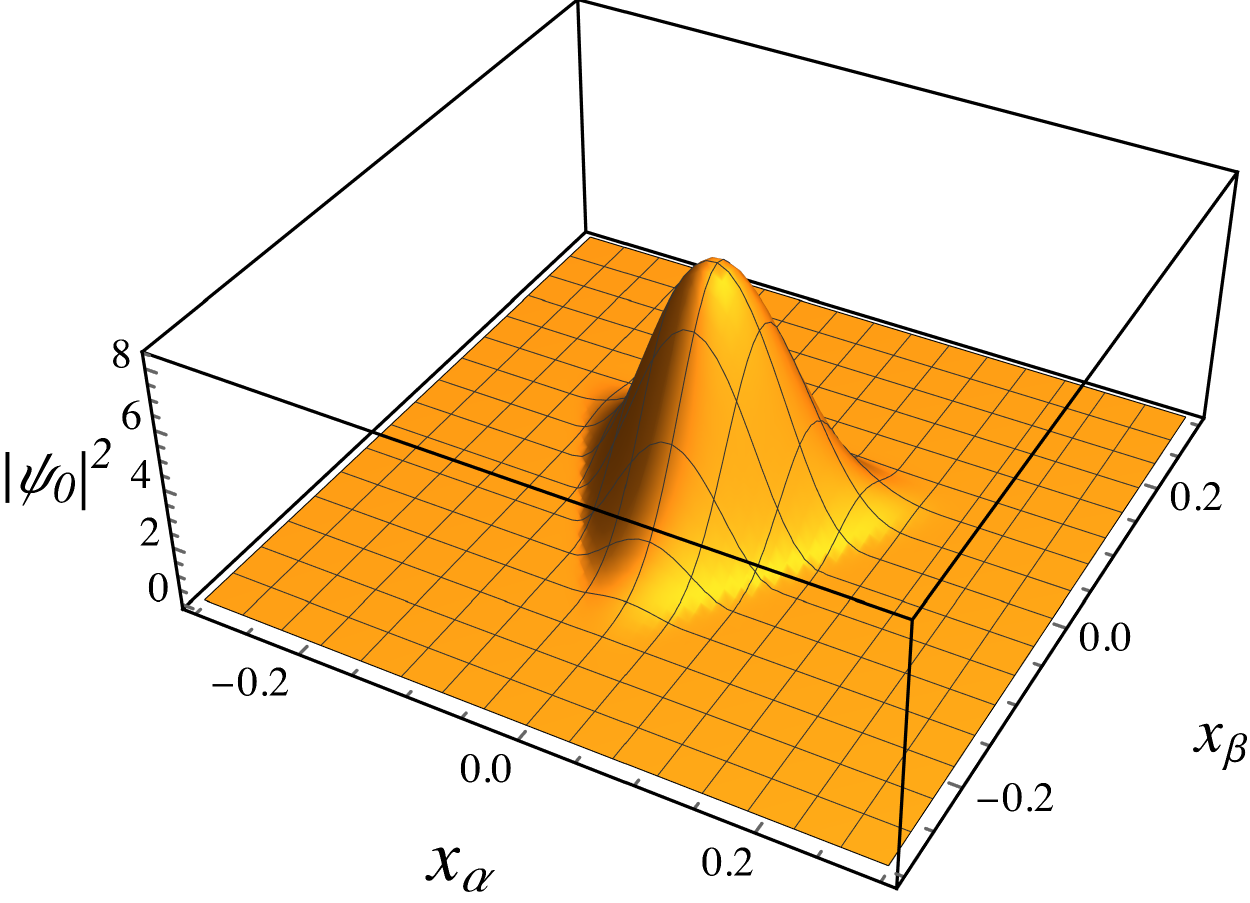} 
\caption{The ground state of the Laplacian operator obtained by diagonalizing the operator using the basis defined in Eq.~(\ref{SLAT4}). The variables $x_\gamma$ and $x_\delta$ are set to $x_\gamma=x_\delta=-0.2$.} 
\label{fig:gl_lap}
\end{center}
\end{figure}

The ground state eigenfunction of the longitudinal Laplacian operator $-\nabla_L$ defined in in Eq.~(\ref{eq:laplacian}) is shown in Fig~\ref{fig:gl_lap} in the $x_\alpha,x_\beta$ plane. Both remaining coordinates, $x_\gamma$ and $x_\delta$, have been set to $x_\gamma=x_\delta=-0.2$.

\subsection{Transverse basis for  N=5}
The LF Hamiltonian in~(\ref{eq:LFHamiJob}) includes 
the non-factorizable part of the potential at the origin of the pentachoron or 
4-simplex 
\ba
V=\sum_{i=1}^5 \vec k_i^2\bigg(\frac 1{x_i}-5\bigg)
\ea
The subtracted mean parton momentum $\frac 15$ centers the longitudinal momenta 
around the center of the 4-simplex. With this in mind, 
the LF Hamiltonian in~(\ref{eq:LFHamiJob}) can be recast as
\bea
H_{LF}\label{eq:Harfina}
&=&\sum_{i=1}^5\frac{m_Q^2}{x_i}-\sum_{\xi=\alpha,\beta,\gamma,\delta}\frac{\sigma_T}{a}(\partial/\partial x_\xi)^2)
\nonumber\\
&+&5\sigma_T a+V
\nonumber\\
&+&\sum_{\xi=\alpha,\beta,\gamma,\delta}-\frac{\sigma_TM^2}{a} \left(\frac{\partial}{\partial_{\vec{k}_{\xi\perp}}}\right)^2+5\vec{k}^2_{\xi\perp}\nonumber\\
\eea
The last contribution amounts to the sum of transverse oscillators in Jacobi coordinates. 
The corresponding eigenfunctions are  generalized Laguerre polynomials, 
\begin{widetext}
\ba  
\label{eq:lagre}
\psi_{n,m}(k_{\xi\perp},\beta) &=&{1\over \sqrt{\pi}} \beta^{1/4} \sqrt{n! \over (n+|m|)!} e^{-\frac{k_{\xi\perp}^2\beta^{1/2}}{2}+i m \phi}  
\,(k_{\xi\perp}  \beta^{1/4})^{|m|}L_n^{|m|}\big( \beta^{1/2} k_{\xi\perp}^2\big)
\ea
\end{widetext}
with $\beta=5a/(\sigma_T M^2)$, and eigenvalues
$$E_{n,m}=2\sqrt{5}(\sigma_T M^2/a)^{1/2}(2n+|m|+1)\,.$$

\begin{table}[h!]
    \centering
    \begin{tabular}{|c|c|c|c|} \hline
  I/S    & 1/2 & 3/2   & 5/2  \\  \hline
  1/2 & 3  & 3    & 1  \\
  3/2 & 3 & 3  &  1   \\
  5/2 & 1   & 1  & 0 \\ \hline
    \end{tabular}
    \caption{Spin-isospin assignments and their degeneracies for the light s-wave pentastates. S and I represent the spin and isospin, respectively.}
    \label{fig:tab1}
\end{table}

\section{Hyperfine interactions}
\label{sec_V}
In addition to the basic light front Hamiltonian used so far ( kinetic 
energy plus confinement), we consider the hyperfine interactions which split the degeneracy of the light pentaquark states, with the spin-isospin assignments shown in Table~\ref{fig:tab1}. As the spin-orbit and tensor interactions vanish in the S-shell, we 
will consider the color-spin exchange expected from perturbative one-gluon exchange, or instanton induced interactions between massive quarks. For comparison with earlier analyses~\cite{Glozman:1995fu,Helminen:2000jb}, we will also carry the analysis for flavor-spin hyperfine interactions, although some
arguments still persist on their importance~\cite{Isgur:1999jv,glozman2000isgurscritiquepionexchange}.

\subsection{Interactions}
With this in mind, the color-spin and flavor-spin hyperfine interactions are respectively
\begin{widetext}
\bea
\label{PERT}
\mathbb V_{1g}(1,2,3,4,5)&=&-\sum_{i<j\leq 5}\frac{V_{1g}(r_{ij})}{m_Q^2}\lambda_{ij}\Sigma_{ij}
\rightarrow -\frac{V_{1g}}{m_Q^2}
\bigg(\sum_{i<j\leq 4}{\lambda_i\cdot\lambda_j\,\sigma_i\cdot\sigma_j}
+\sum_{i<5}{\lambda_i\cdot\bar{\lambda}_5\,\sigma_i\cdot\bar{\sigma}_5}\bigg)   \nonumber\\
\mathbb V_{f}(1,2,3,4,5)&=&-\sum_{i<j\leq 5}C_\chi(r_{ij}) T_{ij}\Sigma_{ij}
\rightarrow -C_\chi
\bigg(\sum_{i<j\leq 4}{\tau_i\cdot\tau_j\,\sigma_i\cdot\sigma_j}
+\sum_{i<5}{\tau_i\cdot\bar{\tau}_5\,\sigma_i\cdot\bar{\sigma}_5}\bigg) 
\eea
\end{widetext}
are sandwiched between the
pertinent color-spin wavefunctions in (\ref{PENTA1}). 
Note that when the exchange
involves the antiquark labeled by 5,  the replacement of the Gell-Mann matrices
$\bar{\lambda}_5=-\lambda^*$ is assumed, and similarly $\bar{\sigma}_5=-\sigma^*$ for the spin below (the iso-spin matrix $\tau$ is also a Pauli matrix). The rightmost result refers to the local approximation  to be used below for a spectrum estimate.

\subsection{Representations of permutation group $S_4$}
In this work we consider the  case of pentaquarks formed of only the light $u,d$ quark flavors of equal mass. The ground state wavefunction carries $L=0$ or space symmetric S-state. The ground state can be constructed using the irreducible representation of the permutation group of four quarks $S_4$. The main results will be summarized below, and more details can be found in the Appendix of Ref.~\cite{Miesch:2025wro}.
More specifically, the totally antisymmetric color-spin-flavor wavefunction in the S-state reads~\cite{Miesch:2025wro}
\begin{widetext}
\bea
\label{PENTA1}
\Psi^S_P&=&\frac {\varphi_{00_L}}{\sqrt 3}\,
\bigg(\bigg[
C[211]_\beta SF[31]_\alpha-C[211]_\alpha SF[31]_\beta+C[211]_\gamma\,SF[31]_\gamma\bigg]C[11] SF[1]\bigg)
\eea
\end{widetext}
where $C[211]$ are the irreducible representation of $S_4$ for the color indices. Their Young tableaux are
\bea
\label{Y2}
C[211]_\alpha=
\begin{ytableau}
1&3\\2\\4\\
\end{ytableau}
\eea
\bea
C[211]_\beta  =
\begin{ytableau}
1&2\\3\\4\\
\end{ytableau}
\eea
\bea
C[211]_\gamma=
\begin{ytableau}
1&4\\2\\3\\
\end{ytableau}\nonumber\\
\eea

For the ground state wavefunction with $L=0$, the 4-quarks spin-flavor mixed representation conjugate to the 4-quarks color representation is also 3-degenerate, with the Jacobi-like nomenclature
\bea
SF[31]_\alpha=
\begin{ytableau}
1&3 & 4\\2\\
\end{ytableau}
\eea
\bea
SF[31]_\beta=
\begin{ytableau}
1&2 & 4\\3\\
\end{ytableau}
\eea
\bea
SF[31]_\gamma=
\begin{ytableau}
1&2 & 3\\4\\
\end{ytableau}
\eea

These mixed representations follow from the product representations 
\bea
\label{PRODUCT}
[q^4]_S\otimes [q^4]_F
\eea
with a mixed $[31]_{S,F}$ net component. The spin flavor configurations of $[q^4]$ follow from
the standard $SU(2)$ Young tableaux, 
\begin{widetext}
\bea
\label{PRODUCT1}
[q^4]_S=
\begin{ytableau}
$$&&&\\
\end{ytableau}
\,\oplus\,
\begin{ytableau}
$$& &\\ \\
\end{ytableau}\,\oplus\,
\begin{ytableau}
$$& \\ &\\
\end{ytableau}\equiv S[4]\oplus  S[31]\oplus  S[22]
\eea
\end{widetext}
and similarly for $[q^4]_F$ with $S\rightarrow F$,
It follows 
from the character representations of $S_4$ that the product representations (\ref{PRODUCT}) using (\ref{PRODUCT1}) are 
\bea
S[4]\otimes F[31]&=&SF[31]\nonumber\\
S[31]\otimes F[4]&=&SF[31]\nonumber\\
S[31]\otimes F[31]&=&SF[4]\oplus SF[31]\oplus SF[22]\nonumber\\
S[31]\otimes F[22]&=&SF[31]\nonumber\\
S[22]\otimes F[22]&=&SF[4]\oplus SF[22]
\eea
 and the spin-flavor exchanged products
\bea
F[4]\otimes S[31]&=&SF[31]\nonumber\\
F[31]\otimes S[31]&=&SF[4]\oplus SF[31]\oplus SF[22]\nonumber\\
F[31]\otimes S[22]&=&SF[31]\nonumber\\
F[31]\otimes S[4]&=&SF[31]
\eea
hence
\begin{widetext}
\bea
SF[31]\subset  S[4]\otimes F[31], S[31]\otimes F[4], S[31]\otimes F[31], S[31]\otimes F[22], S[22]\otimes F[31]
\eea
\end{widetext}

\subsection{Matrix elements}
In Table.~\ref{tab:scmax}, we show the matrix element of color-spin $\lambda_{ij} \Sigma_{ij}$ interaction for  the different pentastates. To distinguish the states with same spin(S), isospin(I) originating from different spin-isospin configurations, we will use the following notations
\begin{widetext}
\bea
S[I]=&&\frac12[\frac12]_\text{a} :  S[31]\otimes F[31],~~~~~~ \frac12[\frac12]_\text{b} :  S[31]\otimes F[22],~~~~~~ \frac12[\frac12]_\text{c} :  S[22]\otimes F[31]  \nonumber\\
&&\frac12[\frac32]_\text{a} :  S[31]\otimes F[4],~~~~~~ \frac12[\frac32]_\text{b} :  S[22] \otimes F[31],~~~~~~ \frac12[\frac32]_\text{c} :  S[31]\otimes   F[31]    \nonumber\\
&&\frac32[\frac12]_\text{a} :  S[4]\otimes F[31],~~~~~~ \frac32[\frac12]_\text{b} :  S[31] \otimes F[22],~~~~~~ \frac32[\frac12]_\text{c} :  S[31]\otimes   F[31]    \nonumber\\
&&\frac32[\frac32]_\text{a} :  S[31]\otimes F[4],~~~~~~ \frac32[\frac32]_\text{b} :  S[4] \otimes F[31],~~~~~~ \frac32[\frac32]_\text{c} :  S[31]\otimes   F[31]    \nonumber\\
&&\frac12[\frac52] : S[31]\otimes F[4],~~~~~~~\frac32[\frac52] : S[31]\otimes F[4],  \nonumber\\
&&\frac52[\frac12] : S[4]\otimes F[31],~~~~~~~\frac52[\frac32] : S[4]\otimes F[31]
\eea
\end{widetext}
On the LF, the quantum states carry fixed $S_z$
in the ground state. On the other hand, the 
spin-color interaction is the same for fixed $S$ but different $S_z$. In Table~\ref{tab:scmax} we show the color-spin interaction for fixed $S_z$ For example, the states with $I=1/2$ and $S_z=1/2$ include the states $\frac12[\frac12]_\text{a}$, $\frac12[\frac12]_\text{b}$, $\frac12[\frac12]_\text{c}$, $\frac32[\frac12]_\text{a}$, $\frac32[\frac12]_\text{b}$, $\frac32[\frac12]_\text{c}$ and $\frac52[\frac12]$. The matrix entries of $\lambda_{ij} \Sigma_{ij}$ between these states 
are
\begin{widetext}
\bea\label{eq:marxs12I12}
\langle S_z=\frac{1}{2},I=\frac{1}{2}|\sum_{i<j\leq5}\lambda_{ij} \Sigma_{ij}|S_z=\frac{1}{2},I=\frac{1}{2}\rangle=\left(\begin{array}{ccccccc} 
-\frac{16}{3} &  0 & 8 & 0 &0 & 0 & 0\\
0 & \frac{56}{3} &0  & 0 &0 & 0 & 0\\
8 &  0 & 0 & 0 &0 & 0 & 0\\
0 &  0 & 0 & 0 &0 & 4\sqrt{10} & 0\\
0 &  0 & 0 & 0 &-\frac{4}{3} & 0 & 0\\
0 &  0 & 0 & 4\sqrt{10} &0 & -\frac{4}{3} & 0\\
0 &  0 & 0 & 0 &0 & 0 & -\frac{40}{3}\\
\end{array}\right)
\eea

In Table~\ref{tab:sfmax} we show the expectations of the spin-flavor interaction  $\sum_{i<j\leq5}\lambda_{ij} \Sigma_{ij}$ in the states with $L=0$, spin $S$ and isospin $I$. The values inside the outer brackets represent the total spin and isospin of the state, respectively. The subscripts a, b, and c indicate different combinations of spin and isospin representations.  The matrix entries of $T_{ij} \Sigma_{ij}$ between these states 
are
\bea\label{eq:marfs12I12}
\langle S_z=\frac{1}{2},I=\frac{1}{2}|\sum_{i<j\leq5}T_{ij} \Sigma_{ij}|S_z=\frac{1}{2},I=\frac{1}{2}\rangle=\left(\begin{array}{ccccccc} 
14 & -4 & -4 & 0 &0 & 0 & 0\\
-4 & 10 & 8  & 0 &0 & 0 & 0\\
-4 &  8 & 10 & 0 &0 & 0 & 0\\
0 &  0 & 0  & 4 & -2\sqrt{10} & -2\sqrt{10} & 0\\
0 &  0 & 0 & -2\sqrt{10} &10 & 2 & 0\\
0 &  0 & 0 & -2\sqrt{10} &2 & 2 & 0\\
0 &  0 & 0 & 0 &0 & 0 & -6\\
\end{array}\right)
\eea
\end{widetext}

\section{Numerical results}\label{sec:numres}
In this section, we will combine the results given above, for
the basis states and color-spin interactions, and 
present the numerical results for the resulting wavefunctions, the eigenstate  of full light cone Hamiltonian. Additionally, we will show the spectrum of light pentaquark state and comparison to the experimental results.
\subsection{Light cone wave function}
To diagonalize the Light front Hamiltonian~(\ref{eq:Harfina}), we choose the eigenbasis as
\bea
\Psi_{\tilde{n},n,m}(x_\xi,k_{\xi\perp})=\varphi_{\tilde{n}}[\tilde{s}]\psi_{n,m}(k_{\xi\perp},\beta)
\eea
where $\xi=\alpha,\beta,\gamma,\delta$ represents the label of Jacobi coordinates,
$\varphi_{\tilde{n}}[\tilde{s}]$ and $\psi_{n,m}(k_{\xi\perp},\beta)$ are given in~(\ref{SLAT4}) and (\ref{eq:lagre}), respectively. The longitudinal and transverse eigen-sets are truncated using the following constraints
\bea
E_L&=&\sum_{i=2}^{5}\tilde n_{i1}^2-\frac 15\bigg(\sum_{i=2}^{5} \tilde n_{i1}\bigg)^2\leq \frac{1184\pi^2}{5}, \nonumber\\
E_T&=&\sum_{i=1}^{4}(2n_i+|m|_i+1)\leq 14
\eea
In the following, we will set $m_i=0$ since we are interested in the S-wave penta-states. The above cutoffs include 134 states and 126 states in the longitudinal and transverse directions, respectively. Eventually, the ground state eigenfunction can be expressed as
\bea
\Phi(x_\xi,k_{\xi\perp}) = \sum_{\tilde{n},n} C_{\tilde{n},n}\Psi_{\tilde{n},n,0},
\eea
where the coefficient $C_{\tilde{n},n}$ can be obtained by diabolizing the Light cone Hamiltonian defined in Eq.~(\ref{eq:LFHamiJob}) using the eigenbasis we defined before.
The constituent in the vacuum at low resolution runs with the momentum, but at zero momentum it is tied to the vacuum condensate~\cite{Kock:2020frx}
\bea
m_Q=\frac{g_S|\langle\overline \psi\psi\rangle|}{N_c}\rightarrow 383\pm 39\, \rm MeV
\eea
The string tension is fixed either from quarkonia spectroscopy or 
from meson Reggion trajectories (by the rho mass) 
\bea
\sigma_T=\frac {m_\rho^2}\pi\rightarrow (0.44\,\rm{GeV})^{2}
\eea

\begin{figure}[htbp]
\begin{center}
\includegraphics[width=7cm, height=5.5cm]{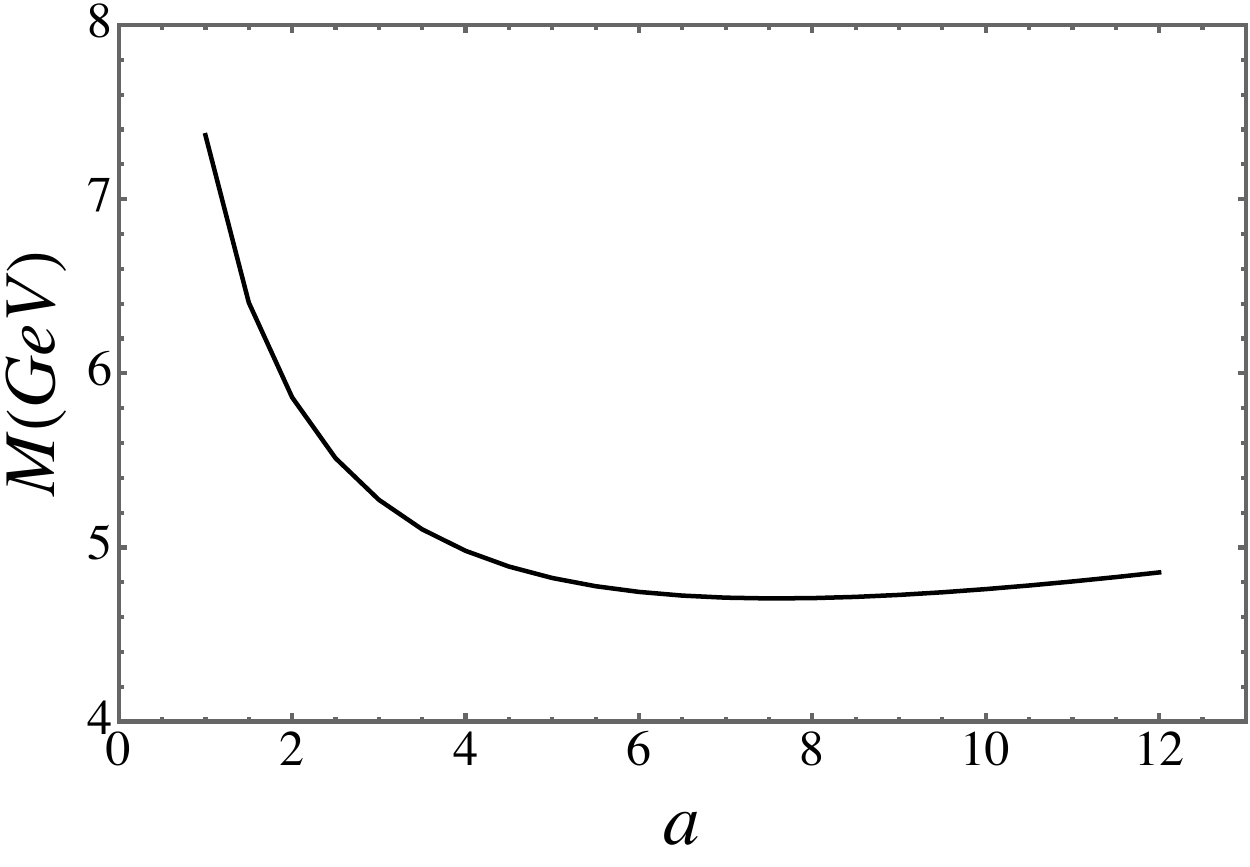} 
\caption{Dependence of ground penta-state mass on the einbein parameter  $a$.} 
\label{fig:massadep}
\end{center}
\end{figure}

In Figure.~\ref{fig:massadep}, we show the dependence of the ground mass on the einbein variable a. The minimal value of the ground state mass sets in around  7.59, which we will use from hereon.

\begin{figure}[H]\label{fig:3dk}
\begin{center}
\includegraphics[width=7cm, height=5.5cm]{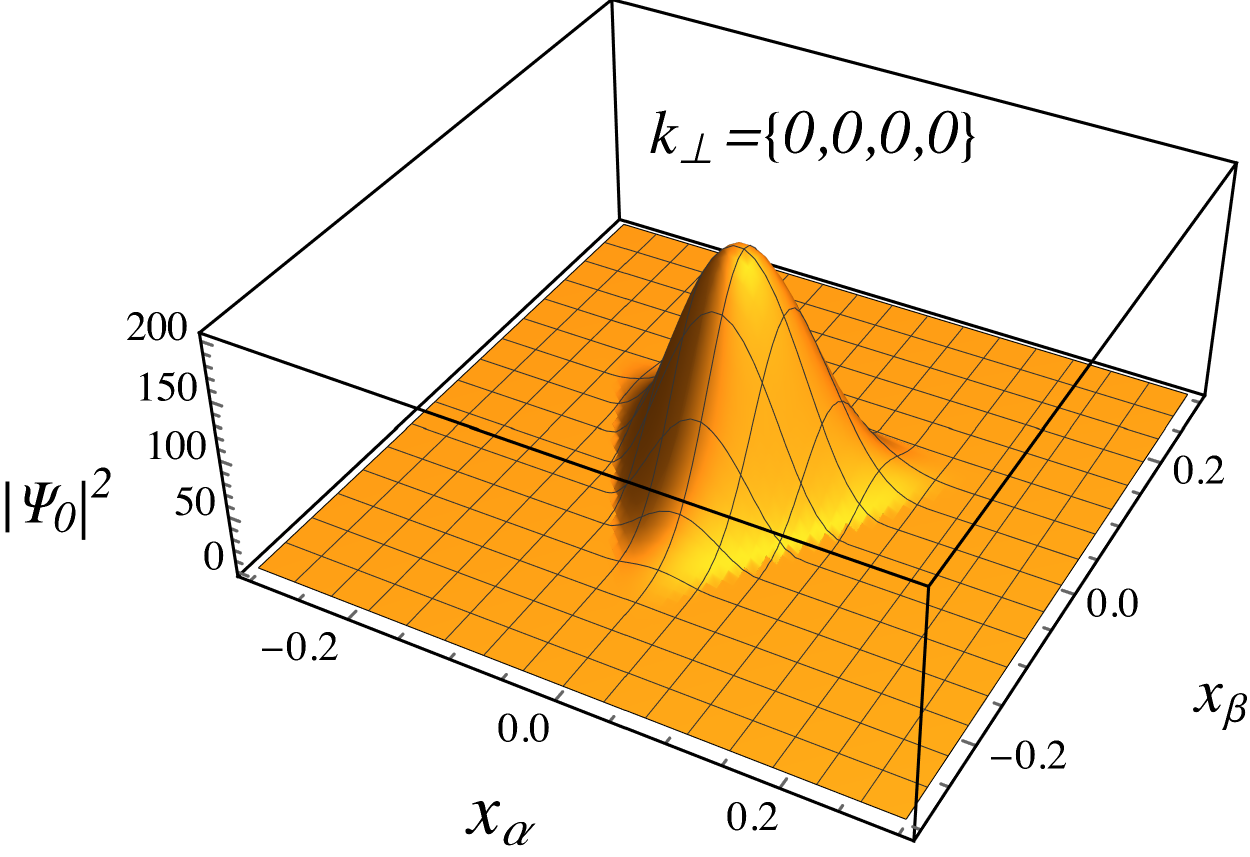} \\
\includegraphics[width=7cm, height=5.5cm]{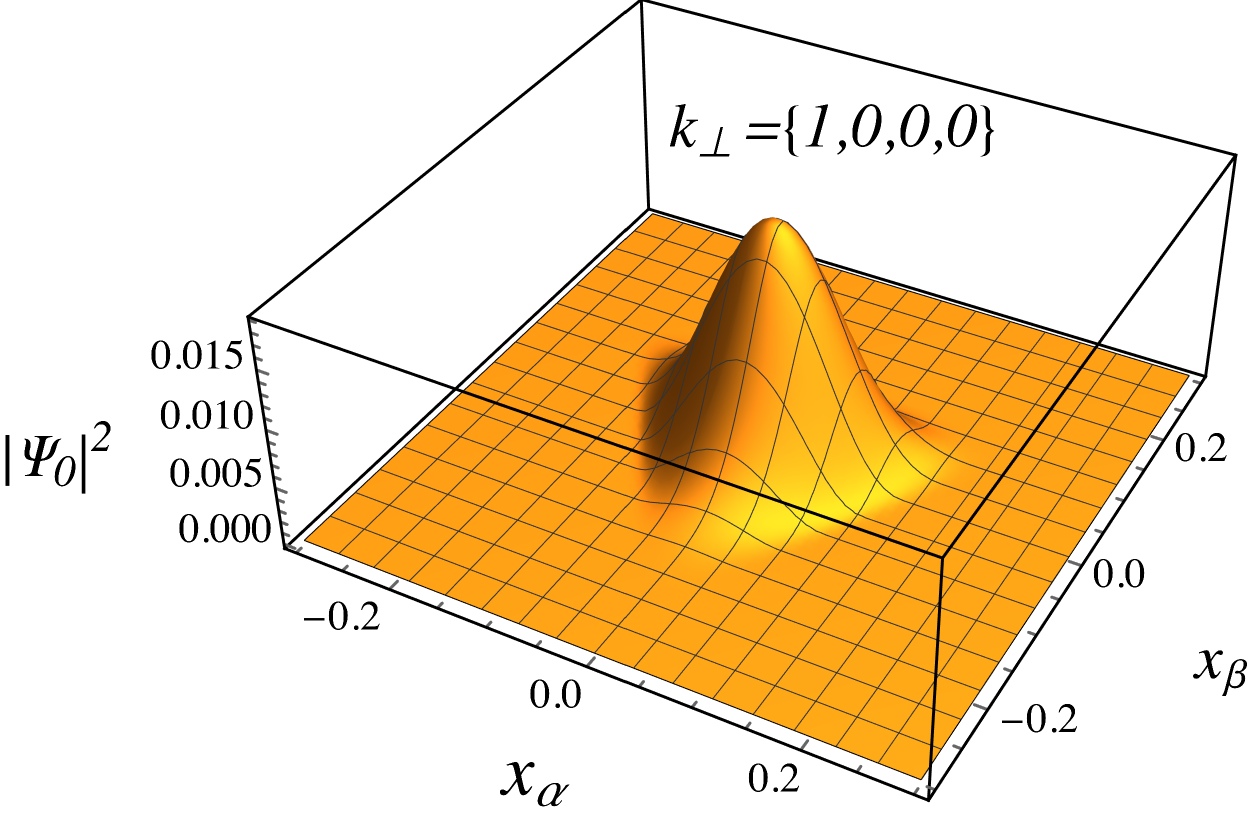} \\
\includegraphics[width=7cm, height=5.5cm]{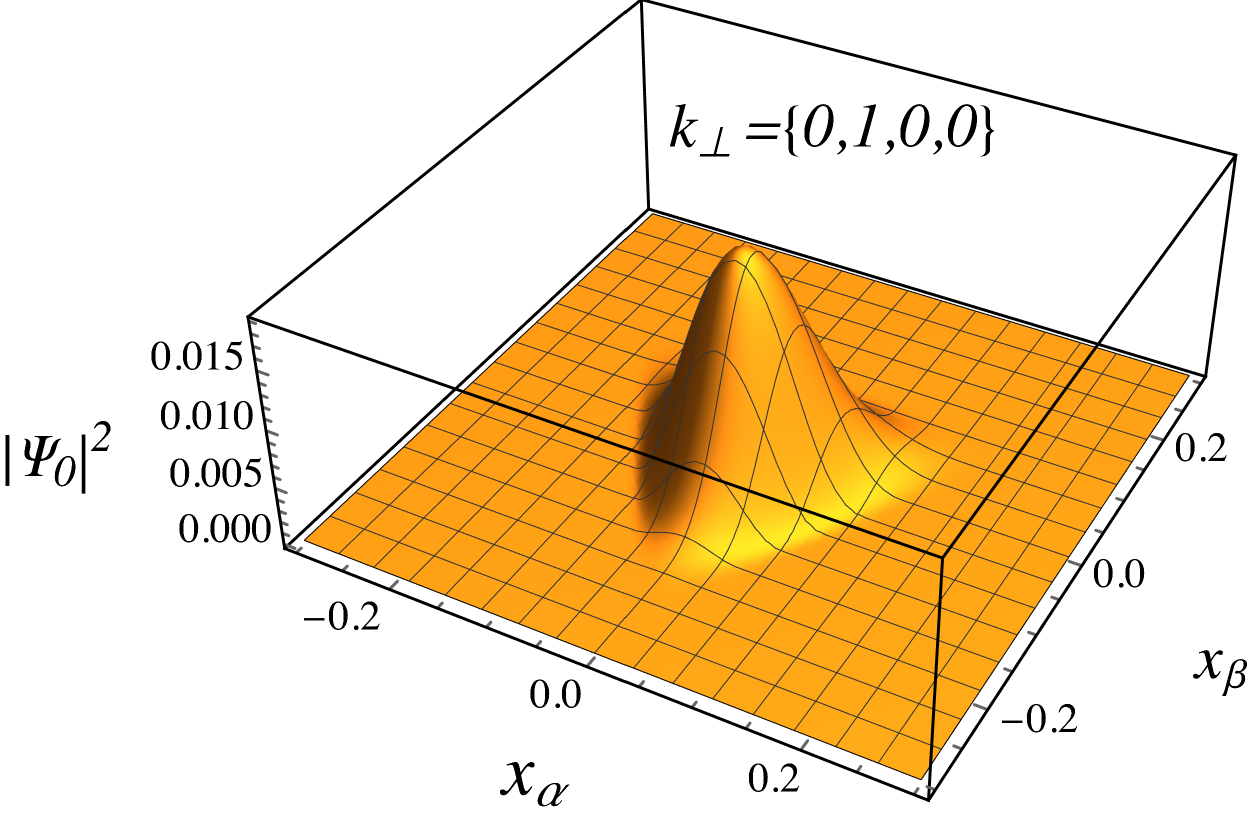} \\
\caption{The ground state wave function squared of full Hamiltonian as a function of two Jacobi coordinates $x_\alpha,x_\beta$. (Other longitudinal variables $x_\gamma$ and $x_\delta$ are set to $x_\gamma=x_\delta=-0.2$.) The corresponding values of the transverse momentum are labeled in the legend as $k_\perp=(|k_\alpha|,|k_\beta|,|k_\gamma|,|k_\delta|).$} 
\label{fig:wf_full}
\end{center}
\end{figure}

In Fig.~\ref{fig:wf_full} we show the longitudinal wavefunction of the ground penta-state obtained by diagonalizing the full Hamiltonian. The longitudinal variables $x_\gamma$ and $x_\delta$ are set to $x_\gamma=x_\delta=-0.2$. The corresponding  Jacobi transverse momenta are labeled as $k_\perp=(|k_\alpha|,|k_\beta|,|k_\gamma|,|k_\delta|)$.

\subsection{Parton distribution function in light pentastates}
\begin{figure}
\begin{center}
\includegraphics[width=7cm, height=5.5cm]{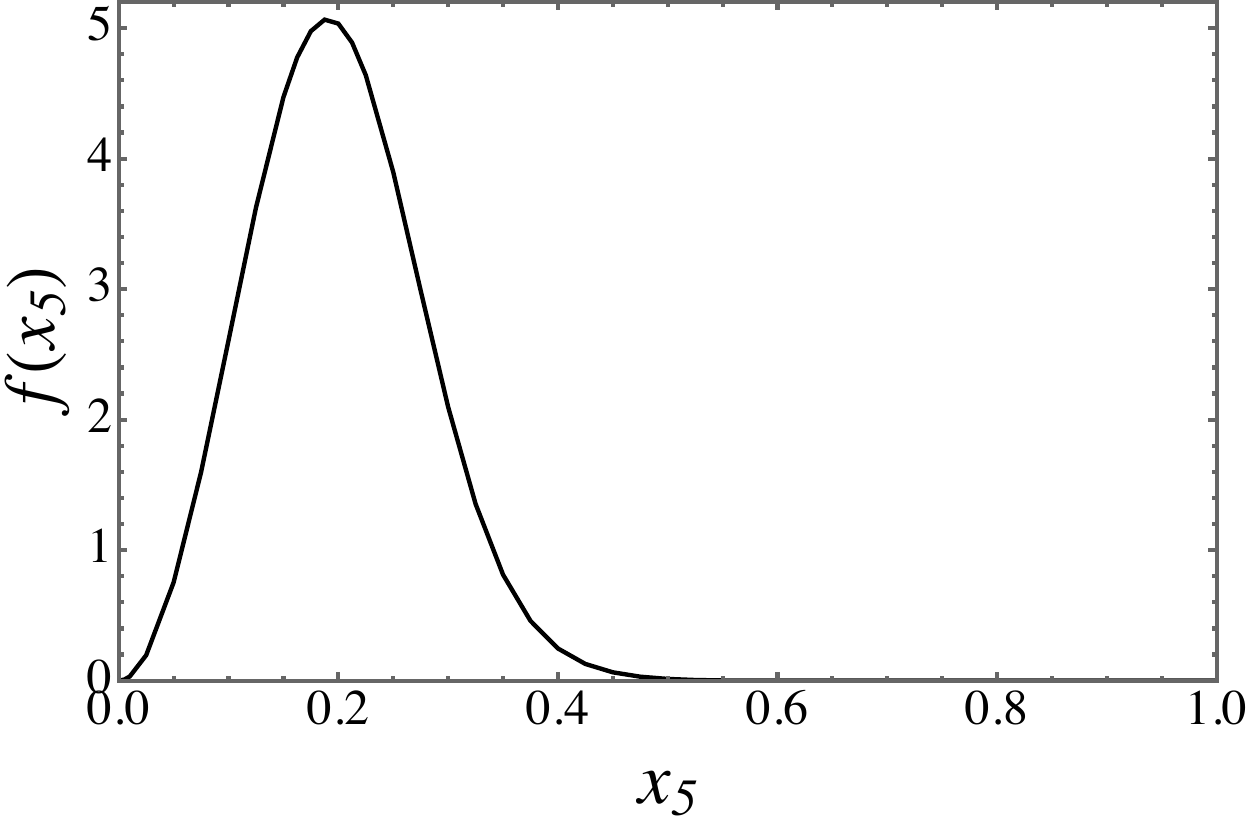} \\
\includegraphics[width=7.2cm, height=5.cm]{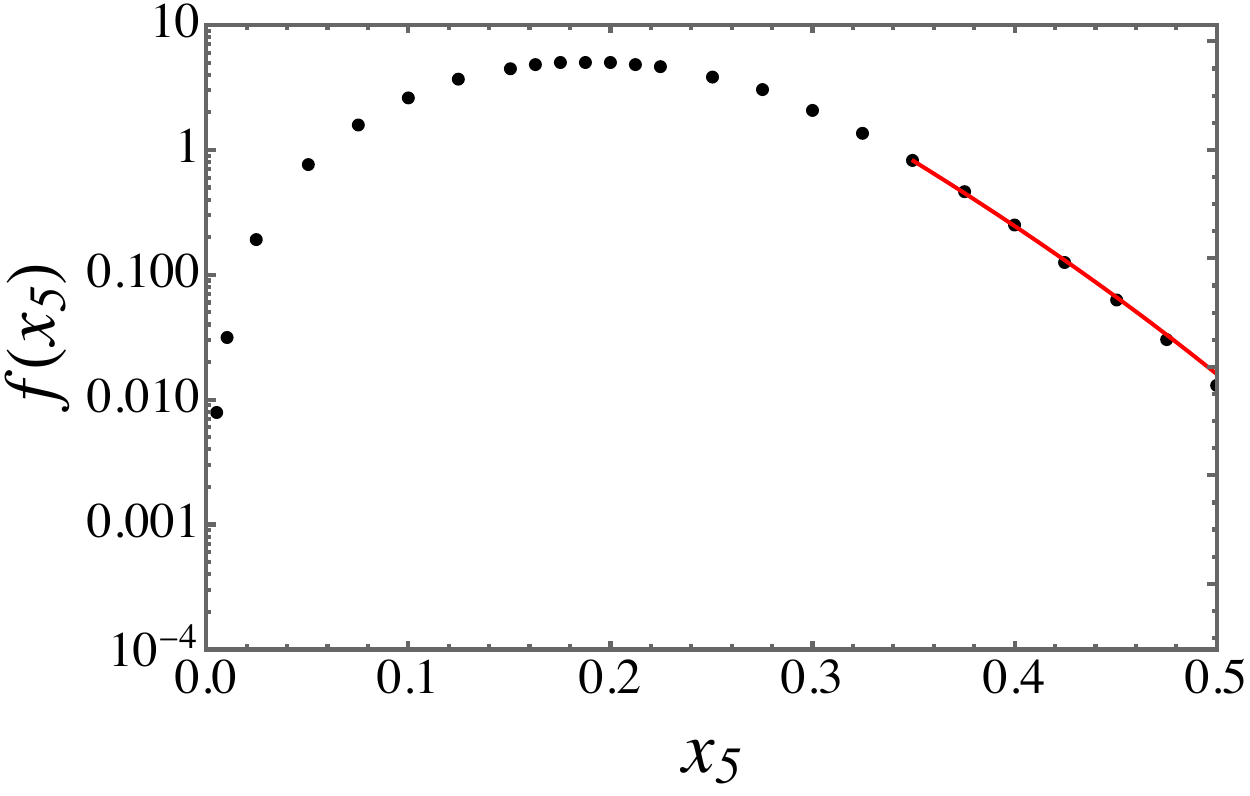}
\caption{The  upper panel shows the quark PDF for S-wave pentastates. The lower panel shows a semi-log plot of the same PDF. The solid red curve shows the asymptotic behavior at large $x_5$, which is given by $(1-x_5)^{15}$.} 
\label{fig:pdf}
\end{center}
\end{figure}

The parton distribution function(PDF) carried by the 5-antiquark inside the pentastate follows from the LF  wave function $\Psi_{\tilde{n},n,m}$,
\bea
&&f(x_5)=\nonumber\\
&&\frac{\sqrt{5}}{2}\int d{x_\xi} d^2 \vec{k}_{\xi\perp}|\Phi({x_\xi,\vec{k}_{\xi\perp}}|^2\delta\left(x_\delta+\frac{5x_5-1}{2\sqrt{5}}\right) \nonumber\\
\eea
where $d{x_\xi}=dx_\alpha dx_\beta dx_\gamma dx_\delta$ and similarly for $d^2 \vec{k}_{\xi\perp}$. The delta-constraint follows from~(\ref{eq:xtoJacb}). The above PDF satisfies the normalization condition $\int_0^1 dx_5 f(x_5)=1$, which follows the LF wavefunction normalization
$$\int d{x_\xi} d^2 \vec{k}_{\xi\perp}|\Phi({x_\xi,\vec{k}_{\xi\perp}}|^2=1\,.$$ 
The quark PDF in the light pentastate  $f(x_5)$ is shown in Fig.~\ref{fig:pdf} (upper panel), with a tail of $(1-x_5)^{15}$ (lower panel). 
Most of the PDF  support is within $0<x<0.6$, with a peak at about $x_5\sim0.2$, consistent with the expectation of equal momentum fraction between the 5 constituents. The PDF at large $x_5$ asymptotes $\sim(1-x_5)^{15}$. Note that for the S-wave analyzed here, there is no difference between the quark and anti-quark PDFs in the pentastate.

\subsection{Light pentastates spectrum}
To estimate the pentastate spectra, we now fix the strengths of the hyperfine interactions. Assuming the  Delta-nucleon mass splitting to be totally due to the color-spin interaction, we have
\bea
M_\Delta-M_N=\frac{16V_{1g}^2}{m_Q^2}\rightarrow (1232-939)\,\rm MeV
\eea
For the spin-flavor interaction we set~\cite{Helminen:2000jb} 
\bea
C_\chi=21\,\text{MeV}
\eea

The pentastate spectrum extracted from the LF analysis is shown in Table~\ref{tab:spec}. The splittings for both the color-spin  interaction (spetrum1) and the spin-flavor interaction (spetrum2), are listed for comparison. The results shown in the second and third column of Table~\ref{tab:spec}, follow from the diagonalization of Eq.~(\ref{eq:marxs12I12}) and Eq.~(\ref{eq:marfs12I12}), respectively. The color-spin and spin-flavor splittings are given in MeV.

The centroids  parameters $c_0$ and $c_1$ are fitted. The reason is that our 
LF Hamiltonian is defined modulo a constant free shift. Indeed, the minimal pentastate  following from our un-shifted LF Hamiltonian in Fig.~\ref{fig:massadep} is about $4.7$ GeV. This high mass value is readily
shifted down by the free and unfixed constant.

To compare with the experimental results, we fix the overall parameter $c_0$ and $c_1$ using the mass of N(1650). The forth and fifth columns show our results for the different states with isospin  $I=\frac{1}{2}$ and spin assignments $\frac 12^-, \frac 32^-, \frac 52^-$. Most of these states are comparable with the empirically listed nucleon states in the PDG~\cite{ParticleDataGroup:2024cfk}. The exception is the  spin $\frac 52^-$, which is 300~MeV heavier than the empirical one.

\begin{table*}[htbp]
    \centering
    \begin{tabular}{|c|c|c|c|c|c|c|c|c|c|c|c|c|c|c|c|c|} \hline
  & $\frac12[\frac12]_\text{a}$ & $\frac12[\frac12]_\text{b}$ & $\frac12[\frac12]_\text{c}$ & $\frac12[\frac32]_\text{a}$ & $\frac12[\frac32]_\text{b}$  & $\frac12[\frac32]_\text{c}$ & $\frac12[\frac52]$ & $\frac32[\frac12]_\text{a}$ & $\frac32[\frac12]_\text{b}$ & $\frac32[\frac12]_\text{c}$ & $\frac32[\frac32]_\text{a}$ & $\frac32[\frac32]_\text{b}$& $\frac32[\frac32]_\text{c}$ & $\frac32[\frac52]$ &$\frac52[\frac12]$ &$\frac52[\frac32]$ \\  \hline
  $\frac12[\frac12]_\text{a}$ & $-\frac{16}{3}$  &     & 8  &  & & & & & &  & & & & & & \\  \hline
  $\frac12[\frac12]_\text{b}$ &   & $\frac{56}{3}$    &   &  & & & & & &  & & & & & & \\  \hline
  $\frac12[\frac12]_\text{c}$ & 8  &     & 0  &  & & & & & &  & & & & & & \\  \hline
  $\frac12[\frac32]_\text{a}$  &   &     &   & $-\frac{88}{3}$ & & & & & &  & & & & & & \\  \hline
  $\frac12[\frac32]_\text{b}$ &   &     &   &  & 0 &8 & & & &  & & & & & & \\  \hline
  $\frac12[\frac32]_\text{c}$ &   &     &   &  &8 & $-\frac{16}{3}$ & & & &  & & & & & & \\  \hline
  $\frac12[\frac52]$ &   &     &   &  & & & $-\frac{88}{3}$ & & &  &  & & & & & \\  \hline
  $\frac32[\frac12]_\text{a}$ &   &     &   &  & & & &  0 & & $4\sqrt{10}$ & & & & & & \\  \hline
  $\frac32[\frac12]_\text{b}$ &   &     &   &  & & & & & $-\frac{4}{3}$ &  & & & & & & \\  \hline
  $\frac32[\frac12]_\text{c}$ &   &     &   &  & & & & $4\sqrt{10}$ & & $-\frac{4}{3}$ &  & & & & & \\  \hline
  $\frac32[\frac32]_\text{a}$ &   &     &   &  & & & & & &  &$-\frac{40}{3}$ & & & & & \\  \hline
  $\frac32[\frac32]_\text{b}$ &   &     &   &  & & & & & &  & & 0 & $4\sqrt{10}$& & & \\  \hline
  $\frac32[\frac32]_\text{c}$ &   &     &   &  & & & & & &  & &$4\sqrt{10}$ &$-\frac{4}{3}$  & & & \\  \hline
  $\frac32[\frac52]$ &   &     &   &  & & & & & &  & & & &$-\frac{40}{3}$ & & \\  \hline
  $\frac52[\frac12]$ &   &     &   &  & & & & & &  & & & & & $-\frac{40}{3}$& \\  \hline
  $\frac52[\frac32]$ &   &     &   &  & & & & & &  & & & & & & $-\frac{40}{3}$\\  \hline
    \end{tabular}
    \caption{The expectational value of $\sum_{i<j\leq5}\lambda_{ij} \Sigma_{ij}$ in the states with $L=0$, spin $S$ and isospin $I$. The number outside and inside the brackets represent the total spin and isospin of the state, respectively. The subscripts a, b, and c indicate different combinations of spin and isospin representations. An empty entry indicates that the corresponding expectation value is zero.}
    \label{tab:scmax}
\end{table*}

\begin{table*}[htbp]
    \centering
    \begin{tabular}{|c|c|c|c|c|c|c|c|c|c|c|c|c|c|c|c|c|} \hline
  & $\frac12[\frac12]_\text{a}$ & $\frac12[\frac12]_\text{b}$ & $\frac12[\frac12]_\text{c}$ & $\frac12[\frac32]_\text{a}$ & $\frac12[\frac32]_\text{b}$  & $\frac12[\frac32]_\text{c}$ & $\frac12[\frac52]$ & $\frac32[\frac12]_\text{a}$ & $\frac32[\frac12]_\text{b}$ & $\frac32[\frac12]_\text{c}$ & $\frac32[\frac32]_\text{a}$ & $\frac32[\frac32]_\text{b}$& $\frac32[\frac32]_\text{c}$ & $\frac32[\frac52]$ &$\frac52[\frac12]$ &$\frac52[\frac32]$ \\  \hline
  $\frac12[\frac12]_\text{a}$ & $14$  &  $-4$   &  $-4$ &  & & & & & &  & & & & & & \\  \hline
  $\frac12[\frac12]_\text{b}$ & $-4$  & 10   & 8  &  & & & & & &  & & & & & & \\  \hline
  $\frac12[\frac12]_\text{c}$ & $-4$  &  8   &  10 &  & & & & & &  & & & & & & \\  \hline
  $\frac12[\frac32]_\text{a}$  &   &     &   & 4 & $-2\sqrt{10}$ & $-2\sqrt{10}$ & & & &  & & & & & & \\  \hline
  $\frac12[\frac32]_\text{b}$ &   &     &   & $-2\sqrt{10}$ & 10 & 2 & & & &  & & & & & & \\  \hline
  $\frac12[\frac32]_\text{c}$ &   &     &   & $-2\sqrt{10}$ &  2  & 2 & & & &  & & & & & & \\  \hline
  $\frac12[\frac52]$ &   &     &   &  & & & $-6$ & & &  &  & & & & & \\  \hline
  $\frac32[\frac12]_\text{a}$ &   &     &   &  & & & &  4 & $-2\sqrt{10}$ & $-2\sqrt{10}$ & & & & & & \\  \hline
  $\frac32[\frac12]_\text{b}$ &   &     &   &  & & & & $-2\sqrt{10}$ & 10 & 2 & & & & & & \\  \hline
  $\frac32[\frac12]_\text{c}$ &   &     &   &  & & & & $-2\sqrt{10}$ & 2 & 2 &  & & & & & \\  \hline
  $\frac32[\frac32]_\text{a}$ &   &     &   &  & & & & & &  &$-5$ & 5 & $\sqrt{10}$& & & \\  \hline
  $\frac32[\frac32]_\text{b}$ &   &     &   &  & & & & & &  & 5 & $-5$ & $\sqrt{10}$& & & \\  \hline
  $\frac32[\frac32]_\text{c}$ &   &     &   &  & & & & & &  &  $\sqrt{10}$ &$\sqrt{10}$ & 8 & & & \\  \hline
  $\frac32[\frac52]$ &   &     &   &  & & & & & &  & & & & 0 & & \\  \hline
  $\frac52[\frac12]$ &   &     &   &  & & & & & &  & & & & & $-6$& \\  \hline
  $\frac52[\frac32]$ &   &     &   &  & & & & & &  & & & & & & 0\\  \hline
    \end{tabular}
    \caption{Similar as Table.~\ref{tab:scmax} but for the expectational value of spin-flavor interaction $\sum_{i<j\leq5}T_{ij} \Sigma_{ij}$.}
    \label{tab:sfmax}
\end{table*}

\begin{table*}[htbp]
    \centering
    \begin{tabular}{|c|c|c|c|c|c|} 
    \hline
    & \makecell{Spectrum1 (color-spin)} & \makecell{Spectrum2 (flavor-spin)} & \makecell{Spectrum1 ($c_0=1756$)} & \makecell{Spectrum2 ($c_1=1860$)}  &Exp  \\ \hline
    \multirow{3}{*}{$\left(\frac{1}{2}\right)^-$}  & $c_0-342$  &   $c_1-462$ & 1414  &1398 &N(1535) \\
      & $c_0-106$ &  $c_1-210$ & {\bf 1650}  &{\bf 1650}   & {\bf N(1650)}  \\
       & $c_0+203$ & $c_1-42$&  1959 & 1818  & N(1895) \\
      \hline
    \multirow{3}{*}{$\left(\frac{3}{2}\right)^-$}  & $c_0-220$    & $c_1-336$ & 1536 & 1524 &N(1520)\\
      & $c_0+24$ & $c_1-84$ & 1780 &  1776 &N(1700)\\
      & $c_0+244$ & $c_1+84$  & 2000 &  1944 &N(1875)\\
       \hline
    \multirow{1}{*}{$\left(\frac{5}{2}\right)^-$}  & $c_0+244$ & $c_1+126$ & 2000 & 1986 & N(1675)\\
   \hline
    \end{tabular}
    \caption{The S-state pentastates with isospin  $I=\frac{1}{2}$ and spin assignments $\frac 12^-, \frac 32^-, \frac 52^-$ and comparison to the experimental result (Exp) from PDG. The Spectrum1 and Spectrum2 represent the hadron mass spectrum obtained by considering spin-color interaction and spin-flavor interactions, respectively. The second and third columns show the energy splitting due to these interactions. The overall parameters $c_0$ and $c_1$ are determined by fixing the mass of {\bf N(1650)}.}
    \label{tab:spec}
\end{table*} 

\section{Conclusions}
The main content of the paper is an explicit construction of eigenbasis for the LF Hamiltonian with confining interactions. It is based on a map
from another simplex corresponding to 5 fermions on the circle,
and expressed by Slater determinants. In terms of Jacobi coordinates
the basis states all satisfy exact Dirichlet boundary conditions on the 
regular 4-simplex or pentachoron. This method can be extended to any N particle states.

This eigenbasis is free from the CM motion both for longitudinal and transverse motion, contrary to other approaches using basis LF Hamiltonians~\cite{Vary:2009gt,Wiecki:2014ola,Jia:2018ary}. 
We further constructed the light pentastates on the LF using Hamiltonian as a matrix in this basis, with subsequent diagonalization.

In addition, we also considered the hyperfine color-spin and flavor-spin interactions. In the absence of hyperfine interactions, there are 16 degenerate  S-pentastates. They are constructed using the representation of the permutation group on $S_4$, to enforce Pauli statistics among the 4 light  quark constituents. 
The degeneracy is lifted in the presence of color-spin or flavor-spin interactions. A generic set of ultra-local hyperfine  interactions, yield 
pentastates that are in fair agreement with empirically reported nucleon excited states carrying the same quantum numbers.

The ground state pentastate leads to antiquark PDF with a peak at $x\approx 0.2$, consistent with the expectation that the five constituents share the LF momentum equally. This PDF displays a hard edge of the form $(1-x)^{15}$ at large $x$.

\label{sec_VI}

\begin{acknowledgments}\noindent
FH is supported by the National Science Foundation under award PHY-2412963. WW  is supported by Tsinghua Scholarship for Overseas Graduate Studies.
ES and IZ are supported by the Office of Science, U.S. Department of Energy under Contract No. DE-FG-88ER40388. This research is also supported in part within the framework of the Quark-Gluon Tomography (QGT) Topical Collaboration, under contract no. DE-SC0023646.
\end{acknowledgments}
\clearpage

\bibliography{pentax}

\end{document}